\documentclass[twocolumn,nofootinbib,showpacs,preprintnumbers,superscriptaddress] {revtex4}

\usepackage{amssymb,amsfonts,amsmath,graphicx,}
\usepackage{color}
\usepackage{enumerate}
\usepackage{braket}
\usepackage{epstopdf}

\newcommand{\dis}[1]{\begin{equation}\begin{split}#1\end{split}\end{equation}}
\newcommand{\etal}{et al.\,}

\newcommand{\bfrac}[2]{{\left(\frac{#1}{#2} \right)  }}
\newcommand{\eq}[1]{Eq.~(\ref{#1})}

\newcommand\tev{\,{\rm TeV}}
\newcommand\gev{\,{\rm GeV}}
\newcommand\mev{\,{\rm MeV}}
\newcommand\kev{\,{\rm keV}}

\newcommand\cm{\,{\rm cm}}
\newcommand\km{\,{\rm km}}

\newcommand\kpc{\,{\rm kpc}}
\newcommand\mpc{\,{\rm Mpc}}

\newcommand\dm{{\rm dm}}
\newcommand\mdm{{M_\dm}}

\begin{document}

\title{Constraining dark matter-neutrino interactions with IceCube-170922A}

\author{Ki-Young Choi}
\email{kiyoungchoi@skku.edu}
 \affiliation{Department of Physics, BK21 Physics Research Division, Institute of Basic Science, Sungkyunkwan University,  2066, Seobu-ro, Jangan-gu, Suwon-si, Gyeong Gi-do, 16419 Korea}

\author{Jongkuk Kim}
\email{jongkukkim@skku.edu}
\affiliation{Department of Physics, BK21 Physics Research Division, Institute of Basic Science, Sungkyunkwan University,  2066, Seobu-ro, Jangan-gu, Suwon-si, Gyeong Gi-do, 16419 Korea}

\author{Carsten Rott}
\email{rott@skku.edu}
\affiliation{Department of Physics, BK21 Physics Research Division, Institute of Basic Science, Sungkyunkwan University,  2066, Seobu-ro, Jangan-gu, Suwon-si, Gyeong Gi-do, 16419 Korea}

\begin{abstract}
Astrophysical neutrinos travel long distances from their sources to the Earth traversing dark matter halos of  clusters of galaxies and that of our own Milky Way.  The interaction of neutrinos with dark matter may affect the flux of neutrinos. The recent multimessenger observation of a high energy neutrino, IceCube-170922A, can give a robust upper bound $\sigma /\mdm \lesssim 5.1\times 10^{-23} \cm^2 /\gev $ on the interaction between neutrino and dark matter at a neutrino energy  of $290 \tev$ allowing 90\% suppression. Combining the constraints from CMB and LSS at different neutrino energies, we can constrain models of dark matter-neutrino interactions. 
\end{abstract}


\preprint{}

\maketitle

{\it Introduction}.\quad
Since neutrinos interact only weakly with matter they can propagate cosmological distances without attenuation and are considered to be ideal messenger particles to uncover the mysteries of distant astrophysical objects. 
The recent discovery of a very high energy neutrino, IceCube-170922A, was followed by multimessenger observations including gamma-ray, X-ray, optical, and radio. Through these accompanying observations,  the source of this 290~TeV neutrino could be identified as a flaring blazar located at a distance of 1421~Mpc~\cite{IceCube:2018dnn}.

New interactions of neutrinos with matter in the Universe may affect the propagation of neutrinos by reducing the flux or changing neutrino flavors~\cite{Smirnov:2004zv,Kile:2013ola}. The nondiagonal or nonuniversal matter potential generated by new interactions modifies the neutrino oscillation behavior and could result in deviation from the present expectations.  Strong constraints can be obtained on nonstandard interactions from atmospheric data~\cite{Mitsuka:2011ty}, at the production, propagation and detection~\cite{Rasmussen:2017ert}, and from neutrino experiments~\cite{Farzan:2017xzy}.

Neutrinos could have interactions with dark matter and observations of distant sources are ideal to probe such processes. Dark matter composes 26\% of the mass-energy content of the present Universe and spreads all over the Universe, with more localization near galaxies and clusters of galaxies. Even though the simplest cosmological $\Lambda_{\rm CDM}$ model assumes only gravitationally interacting dark matter, many models of particles physics predict nongravitational interactions of dark matter with standard model particles as well as self interaction between dark matter~\cite{Baer:2014eja}.

The interaction of neutrinos with dark matter, denoted DM, has been considered in cosmology and neutrino observations.
Before the last scattering of CMB, the interactions of DM beyond gravity leads to a suppression of the primordial density fluctuations, and thus erase the small scale structures and suppress the CMB spectrum at small scales~\cite{Boehm:2000gq,Mangano:2006mp,Serra:2009uu,Wilkinson:2014ksa,Bertoni:2014mva,Escudero:2015yka,DiValentino:2017oaw,Diacoumis:2018ezi}.
 
In the present Universe, the interaction of neutrinos with DM can dissipate neutrinos and hence suppress the flux of neutrinos at Earth.
This attenuation  once was considered to explain the suppression of high-energy neutrino flux~\cite{Barranco:2010xt}. 
This suppression also can be used to constrain the interaction of neutrinos and DM, especially for ultralight scalar dark matter~\cite{Barranco:2010xt,Reynoso:2016hjr}. 

Arguelles \etal~\cite{Arguelles:2017atb} considered the present-day interactions between high-energy cosmic neutrinos and the DM halo of the Milky Way. By taking the isotropic distribution of 53 high-energy neutrinos they could constrain DM-neutrino interactions, since the attenuation of the neutrino flux depends on the direction of the source and leads to the energy-dependent anisotropy. 

Pandey \etal~\cite{Pandey:2018wvh} instead considered the significant flux suppression of high-energy astrophysical neutrinos due to the   interactions with dark matter. They allowed 1\% suppression by just assuming the traveling distance of the neutrino as 200~~Mpc  and the cosmological DM density. With other collider search limits, they studied several effective operators for the interaction.

For a long-range interaction about the astrophysical size, the matter effects are integrated over the interaction size and may affect neutrino flavor oscillations. The neutrino flavor distribution at Earth~\cite{Aartsen:2015knd} can constrain the lepton-number symmetries~\cite{deSalas:2016svi,Chao:2017emq,Esteban:2018ppq,Bustamante:2018mzu,Heeck:2018nzc,Huang:2018cwo}.

In this article,  we consider the recent observation of the high energy neutrino, IceCube-170922A, to obtain a robust bound on the interaction of neutrinos with DM at high energy and combine our result with other bounds at different energies.  As a specific example, we use a model of scalar DM with a fermion mediation. \\

{\it multimessenger high energy neutrino: IceCube-170922A}.\quad
A 290~TeV muon neutrino observed on September 22, 2017 and publicized via alert, IceCube-170922A, is the first high energy neutrino whose origin can be identified with high confidence. Its source, the $\gamma$-ray blazar TXS 0506+056,  was located at redshift $z=0.3365\pm0.0010$~\cite{Paiano:2018qeq}, corresponding to a distance $1421^{+4}_{-5}\mpc$, and was established through multimessenger observations~\cite{IceCube:2018dnn} and archival neutrino data analyses~\cite{IceCube:2018cha}. While blazars have long been suggested as sources of astrophysical neutrinos, a recent study concluded that they contribute not more than 27\%~\cite{Aartsen:2016lir} of the observed IceCube astrophysical neutrino flux~\cite{Aartsen:2014gkd,Aartsen:2013jdh}. Given the observation of IceCube-170922A, we can for the first time study the propagation of a high energy neutrino with a known path and distance~\cite{Kelly:2018tyg}.

If neutrinos interact with dark matter, they can undergo dissipation during the propagation and may not arrive at Earth. The dissipation depends on the scattering cross section, $\sigma$, and the dark matter number density, $n$, along the path of the neutrino resulting in a suppression factor given by $\exp(-\int n \sigma dl)$. When the integration in the exponent $\int n \sigma dl$ is much larger than 1, the neutrino flux is exponentially suppressed and becomes unobservable at Earth.

\begin{table}
\begin{center}
\begin{tabular}{|c| c| c|}
\hline
 Neutrino energy &  $\sigma/\mdm [\cm^2/\gev]$ & Exp. [Ref.] \\ 
 \hline \hline
$\sim 100$ eV &$6\times 10^{-31}$& CMB~\cite{Escudero:2015yka,DiValentino:2017oaw,Diacoumis:2018ezi}\\
$\sim 100$ eV &$10^{-33}$& Lyman-$\alpha$~\cite{Wilkinson:2014ksa}\\
$ 10$ MeV &$10^{-22}$& SN1987A~\cite{Mangano:2006mp}\\
290 TeV&$5.1\times10^{-23}$&IceCube-170922A~\cite{IceCube:2018dnn}\\
 \hline
\end{tabular}
 \caption{ Upper bound on the neutrino-DM scattering cross section from different experiments. In the first column, we specified the corresponding neutrino energy for which  each experimental constraint is applied.}
\label{UpperBound}
\end{center}
\end{table}

Since the number density of dark matter may change with propagation, we can approximate the suppression factor as one from the cosmological dark matter and the other from dark matter in our Milky Way
\dis{
\int_{\rm path} \sigma n({\bf x}) dl &=\int_{los} n(z) \sigma dl + \int_{los} \sigma n_{\rm gal}({\bf x}) d l, \\
&= \frac{\sigma}{\mdm} \left( \int_{los}\rho(z) dl + \int_{los}  \rho_{\rm gal} ({\bf x})d l \right).
}
Here $L$ is the distance from the neutrino source to the Earth and $n_0$ and $n_{\rm dm}({\bf x})$ are the DM number densities in the large scale Universe and in the Milky Way. In the second term, we used the relation between DM energy density and DM mass, $\rho_{\rm dm} = n_
\dm \mdm$, to convert the number density to energy density.
We assume that the cosmological DM density, $\rho(z)=\rho_0 (1+z)^3$ with $\rho_0\simeq 1.3\times 10^{-6}\gev/\cm^3$, which is the dark matter density along the path. The DM density in our Milky Way is position dependent and we assume the NFW profile~\cite{Navarro:1996gj} given by
\dis{
\rho_{\rm gal} ({\bf x})= \frac{ \rho_{s}}{ \frac{r}{r_s}\left(1+\frac{r}{r_s} \right)^2},
}
where $\rho_s= 0.184 \gev/\cm^3$, $r_s=24.42 \kpc$ with $\rho_\odot=0.3\gev/\cm^3$, and $r$ is the distance from the Galactic center.

For the neutrino from  IceCube-170922A with the distance $L=1421\mpc$, we find that the cosmological suppression factor is
\dis{
 \int_{los}\rho (z)\,  dl =& \int  \rho(z)  \frac{cdt}{dz} dz,\\
\simeq &\, 7.2\times 10^{21}\gev/\cm^2,
\label{cos}
}
where $dt/dz=-((1+z)H(z))^{-1}$ and $H(z)=H_0\sqrt{\Omega_\Lambda + \Omega_m(1+z)^3}$. The last term was obtained using the present value $H_0 = 67.4 \km/\sec/\mpc$, $\Omega_\Lambda = 0.685$, and $\Omega_m=0.315$~\cite{Aghanim:2018eyx}.

For the suppression due to the DM interaction in the Milky Way, we need to consider the direction of the neutrino source and integrate the number density  along the path of the neutrinos. We find that the suppression factor is
\dis{
 \int_{los}  \rho_{\rm gal} ({\bf x})d l\simeq  3.8 \times 10^{22} \gev/\cm^2.
 \label{MW}
}
For this calculation we use the well known direction of IceCube-170922A  in the right ascension (RA) $77.42^{+0.95}_{-0.65}$ and declination (Dec) $+5.72^{+0.50}_{-0.30}$ to convert it to the Galactic coordinates used in $\rho_{\rm dm}({\bf x})$ of the Milky Way halo.
We find that this result does not depend on the choice of DM halo profile, since the direction to the IceCube-170922A is not the center of the Milky Way from the Earth.
\begin{figure}[!t]
\begin{center}
\begin{tabular}{c} 
 \includegraphics[width=0.4\textwidth]{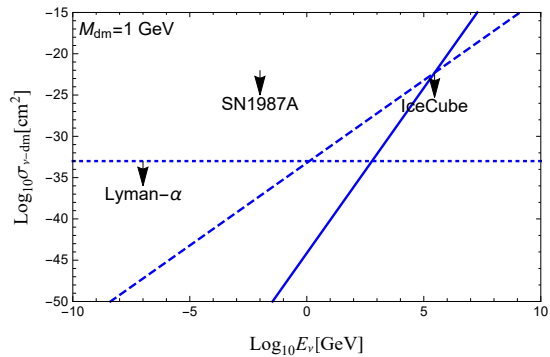}
  \end{tabular}
\end{center}
\caption{Upper bound on the scattering cross section for different energy dependence of scattering of neutrinos with dark matter.  The points of ``IceCube" and  ``Lyman-$\alpha$" are the experimental upper bounds on the cross section for $\mdm=1\gev$ at the corresponding neutrino energy.  Here we used the power-law form $\sigma (E_\nu) = \sigma_0 \bfrac{E_\nu}{1\gev}^n$, with index $n=0,2,4$ for dotted, dashed, and solid lines respectively.}
\label{sigEnu_n}
\end{figure}

\begin{figure}[!t]
\begin{center}
\begin{tabular}{cc} 
 \includegraphics[width=0.4\textwidth]{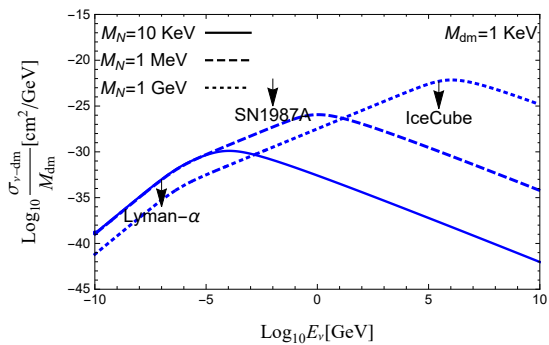}
    \end{tabular}
\end{center}
\caption{The scattering cross section versus neutrino energy for the model of complex scalar DM with a fermion mediation~\cite{Escudero:2015yka}. Here we fixed  $\mdm=1\kev$ and used $m_N=10\kev, 1\mev$, and  $1\gev$, and show the biggest cross section that satisfies the experimental bounds.}
\label{SdmFmed}
\end{figure}

\begin{figure*}[!t]
\begin{center}
\begin{tabular}{cc} 
 \includegraphics[width=0.41\textwidth]{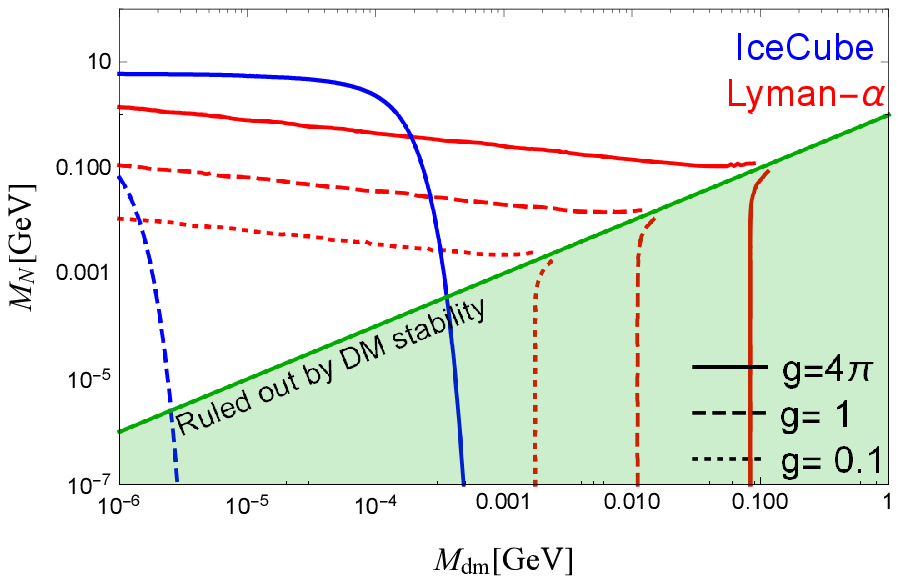}
 &
 \includegraphics[width=0.4\textwidth]{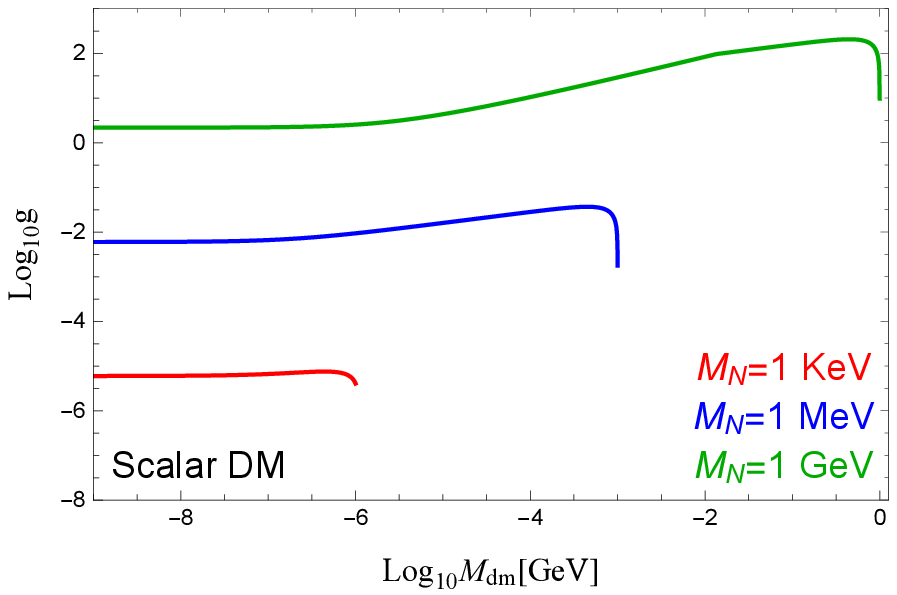}
   \end{tabular}
\end{center}
\caption{Left: The contour plot in the ($\mdm,M_N$) plane which touches the constraint Lyman-$\alpha$ (Red) or IceCube (Blue) for  given couplings $g=0.1, 1$, and $4\pi$. The upper and right region of the line for given coupling is allowed. In the green region DM is heavier than the fermion and thus is not stable. Right: The maximum values of the coupling $g$ versus DM mass for given fermion mass $m_N=1\kev, 1\mev$, and  $1\gev$.}
\label{SdmFmed_con}
\end{figure*}
Incidentally both contributions from cosmological DM and Milky Way DM are very comparable, since the small cosmological DM density is compensated by the long distance.
The observation of the high energy neutrino  IceCube-170922A implies that the neutrino flux did not have significant suppression during its propagation. This enables us to place an upper bound on the interaction of neutrinos with dark matter.
Considering that the suppression is not  larger than 90\% of the original flux, we require $\int \sigma n dl  \lesssim 2.3$.
Using \eq{cos} and \eq{MW}, we can find the upper bound on the scattering cross section as
\dis{
\frac{\sigma }{\mdm} \lesssim & \, 2.3 \times \left(\rho_{0}  L + \int_{los}  \rho_{\rm gal} ({\bf x})d l \right)^{-1} \\
 \simeq &\, 5.1\times 10^{-23} \cm^2 /\gev \quad {\rm at}  \quad E_\nu = 290\tev.
}
assuming that the scattering cross section does not change during the propagation.\\

{\it Upper bound on the neutrino-DM interaction at different energies}.\quad
The present bound on the scattering cross section between neutrinos and DM is summarized in Table~\ref{UpperBound}.
The constraint from CMB and Lyman-$\alpha$ comes from the small scale suppression of the density fluctuation that has been caused before the last scattering of photons, when the neutrino energy was around 100~eV.
Our constraint from IceCube-170922A is applied for a neutrino energy of 290~TeV.\\

{\it Model of simple power-law}.\quad
As the scattering cross section could be energy dependent, we explore simple power-law forms of the energy dependence with $n=0,2,4$ as
\dis{
\sigma(E_\nu) = \sigma_0 \bfrac{E_\nu}{1\gev}^n,
}
where $\sigma_0$ is the cross section normalized at the neutrino energy at $E_\nu=1\gev$.
In Fig.~\ref{sigEnu_n}, we show the constraints on the scattering cross section for different energy dependence with $n=0,2,4$.
For each case, we find the upper bound on $\sigma_0$ as
\dis{
&\sigma_0/\mdm \lesssim 10^{-33} \cm^2/\gev \quad {\rm for} \quad n=0,\\
&\sigma_0/\mdm \lesssim 6.3\times 10^{-34}\cm^2/\gev\quad {\rm for} \quad n=2,\\
&\sigma_0/\mdm \lesssim 7.5\times 10^{-45}\cm^2/\gev\quad {\rm for} \quad n=4.\\
}

{\it Model of complex scalar DM mediated by a  fermion}.\quad
For complex scalar DM with a fermionic mediator, the interaction Lagrangian will be
\dis{
{\mathcal L}_{\rm int} = -g \chi \overline{N}\nu_L + {\rm h.c.},
}
where $g$ is the coupling for the Yukawa interaction between complex dark matter $\chi$,  fermion $N_R$, and left-handed neutrino $\nu_L$. In this case, the mass of DM needs to be smaller than that of the fermion for stable DM. The scattering cross section has nontrivial dependence on the masses and neutrino energy. The cross section scales as $\sigma \propto E_\nu^2$  for $E_\nu \lesssim \mdm$, $\sigma \propto E_\nu$ for $\mdm\lesssim E_\nu \lesssim m_N^2/(2\mdm)$, and $\sigma \propto E_\nu^{-1}$ for $ E_\nu \gtrsim m_N^2/(2\mdm)$.

In Fig.~\ref{SdmFmed}, we show the scattering cross section versus neutrino energy for this  model~\cite{Escudero:2015yka}. Here we fixed  $\mdm=1\kev$ and used $m_N=10\kev, 1\mev$, and  $1\gev$, and show the behavior of the cross section with the biggest coupling that satisfies the experimental bounds in Table~\ref{UpperBound}. 

In Fig.~\ref{SdmFmed_con} (Left), we show the contour plot in the ($\mdm,M_N$) plane which touches the constraint Lyman-$\alpha$ {(Red) or IceCube (Blue) for  given couplings $g=0.1, 1$, and $4\pi$. In the green region DM is heavier than the fermion and thus is not stable.
For a given coupling, in the upper and right region both the blue and red lines are allowed, since the
 strongest bound depends on the neutrino energy.
In Fig.~~\ref{SdmFmed_con} (Right), the upper bound on the coupling is shown versus DM mass for the given mediator mass with $m_N=1\kev, 1\mev$, and  $1\gev$.\\

{\it Conclusion}.\quad
The multimessenger observation of IceCube-170922A identified the source of the neutrino at energy 290~TeV, with the definite distance and direction. With this information we can calculate the precise suppression of the neutrino flux when there is interaction with dark matter in our Milky Way and in the Universe. By allowing a 90\% suppression of the neutrino flux, we derived an upper bound on the neutrino-dark matter scattering cross section as $\sigma /\mdm \lesssim 5.1\times 10^{-23} \cm^2 /\gev $ at the corresponding neutrino energy. Since the scattering cross section depends on the neutrino energy we need to combine the experimental constraints at different energies together to constrain specific microphysics models.

{\it Acknowledgments}.
K.-Y.C. was supported by the National Research Foundation of Korea(NRF) grant funded by the Korea government(MEST) (NRF-2016R1A2B4012302). C.~Rott acknowledges support from the National Research Foundation of Korea (NRF) for the Basic Science Research Program (NRF-2017R1A2B2003666).


\end{document}